\documentclass[useAMS,usenatbib]{mn2e}

\usepackage{graphicx}
\usepackage{txfonts}

\title[Abundances in six spiral galaxies]
      {Oxygen and nitrogen abundances of H\,{\sc ii} regions in six spiral galaxies}

\author[A.S.~Gusev et al.]
       {A.S.~Gusev,$^{1}$
        L.S.~Pilyugin,$^{2}$
        F.~Sakhibov,$^{3}$
        S.N.~Dodonov,$^{4}$
        O.V.~Ezhkova,$^{1}$
\newauthor
        M.S.~Khramtsova$^{5}$ \\
 $^{1}$ Sternberg Astronomical Institute, Lomonosov Moscow State University,
            Universitetsky pr. 13, 119992 Moscow, Russia \\
 $^{2}$ Main Astronomical Observatory of National Academy of Sciences of Ukraine,
            Zabolotnogo str. 27, 03680 Kiev, Ukraine, {\sf pilyugin@mao.kiev.ua} \\
 $^{3}$ University of Applied Sciences of Mittelhessen, Campus Friedberg,
        Department of Mathematics, Natural Sciences and Data Processing, \\
        Wilhelm-Leuschner-Strasse 13, 61169 Friedberg, Germany \\
 $^{4}$ Special Astrophysical Observatory, Russian Academy of Sciences, 
        369169 Nizhnij Arkhyz, Russia \\
 $^{5}$ Institute of Astronomy, Russian Academy of Sciences,
        ul. Pyatnitskaya 48, 119017 Moscow, Russia \\
             }

\date{Accepted 2012 May 15. Received 2012 May 14; in original 
form 2012 February 29}

\begin{document}

\maketitle

\begin{abstract}
Spectroscopic observations of 63 H\,{\sc ii} regions in six spiral galaxies 
(NGC~628, NGC~783, NGC~2336, NGC~6217, NGC~7331, and NGC~7678) were carried out 
with the 6-meter telescope (BTA) of Russian Special Astrophysical Observatory 
with the Spectral Camera attached to the focal reducer SCORPIO in the multislit 
mode with a dispersion of 2.1\AA/pixel and a spectral resolution of 10\AA. 
These observations were  used to estimate the oxygen and nitrogen abundances 
and the electron temperatures in H\,{\sc ii} regions through the recent 
variant of the strong line method (NS calibration). 
The parameters of the radial distribution (the extrapolated central intercept 
value and the gradient) of the oxygen and nitrogen abundances in the disks of 
spiral galaxies NGC~628, NGC~783, NGC~2336, NGC~7331, and NGC~7678 have been 
determined. The abundances in the NGC~783, NGC~2336, NGC~6217, and NGC~7678 
are measured for the first time. Galaxies from our sample follow well the 
general trend in the luminosity -- 
central metallicity diagram for spiral and irregular galaxies.
\end{abstract}

\section{Introduction}

We have observed emission line spectra of 63 giant H\,{\sc ii} regions 
in six spiral  galaxies as a part of our study of star formation regions  
 in spiral and irregular galaxies. 
Giant extragalactic H\,{\sc ii} regions are the birth places of star clusters and can be used to study 
the current star formation and chemical abundances in galaxies.
Giant H\,{\sc ii} regions are ionised by clusters of young massive stars.  
Their sizes range from several tens 
to $\approx 500$ pc, so they are larger and brighter in comparison to Galactic H\,{\sc ii} regions.
Certain selection effects must be noted. Necessarily poorer spatial resolution
 contributes to a tendency to identify larger regions in more distant galaxies;
 at better resolution these regions break up into groups or chains 
of smaller clumps \citep{dinerstein1990}. 

One of the main interests in H\,{\sc ii} regions is in the study of elemental 
abundances and their gradients in galaxies. A large number of regions have been observed 
for this purpose \citep[][among others]{zaritsky1994,roy1996,vanzee1998,dutil1999,
kennicutt2003,bresolin2005,bresolin2009}. 
Radial distributions of oxygen and nitrogen abundances across the disks 
are mandatory in investigations of different aspects of formation 
and evolution of spiral galaxies. 
The measurement of the distribution of elemental abundances within galaxies is a 
tool for studying galaxy formation and evolution. There are several investigations 
of possible relationship between the abundances properties and global characteristics 
of galaxies such a Hubble type and luminosity 
\citep{pagel1991, vilacostas1992, dutil1999, garnett1987}.
The oxygen abundance in the interstellar gas is usually used as a
tracer of metallicity in late type (spiral and irregular) galaxies
at the current epoch.
The study of abundance gradients in the disk of 
spiral galaxies has been  started by \citet{searle1971}, with the recognition of a 
radial abundance gradient in M33. 
Results of investigations of variations in the gas composition within 
galaxies combined with results on the evolution of stellar populations 
provide the development of chemical evolution models
\citep{chiappini2003,marcon2010}. 

A study of abundances and their gradients is based 
on measurements of emission line spectra 
of individual H\,{\sc ii} regions in nearby galaxies. 
To define the parameters of radial distribution of oxygen and 
nitrogen abundances (the extrapolated central intercept value and the gradient),
the abundance measurements for a sufficiently large number of H\,{\sc ii}
regions evenly distributed across the galaxy disk are necessary.
Measurements of this kind are available for the limited 
number ($\sim$50) of nearby galaxies \citep[see compilations in][]{garnett2002,pilyugin2004,moustakas2010}.

\begin{table*}
\caption[]{\label{table:sample}
The galaxy sample.
}
\begin{center}
\begin{tabular}{cccccccccccc} \hline \hline
Galaxy & Type & $B_t$ & RA$^a$ & DEC$^a$ & Inclination & P.A. & $v^b$ & $R_{25}^c$ & $R_{25}^c$ & $d$
& $M_B$ \\
          &          & (mag) & (J2000.0) & (J2000.0) & (degree) & (degree) &  (km/s) & (arcmin) & (kpc) &
(Mpc)     & (mag) \\
   1 & 2 & 3 & 4 & 5 & 6 & 7 & 8 & 9 & 10 & 11 & 12 \\
\hline
NGC~628   & Sc & 9.70 & 01 36 41.81 & +15 47 00.3 & 7 & 25 & 659 & 5.23 & 10.96 & 7.2 & -20.72 \\
NGC~783   & Sc & 13.18 & 02 01 06.59  & +31 52 56.2 & 43 & 57 & 5192 & 0.71 & 14.56 & 70.5 & -22.01 \\
NGC~2336  & SB(R)bc & 11.19 & 07 27 03.98 & +80 10 41.1 & 55 & 175 & 2202 & 2.51 & 23.51 & 32.2 & -22.14 \\
NGC~6217  & SB(R)bc & 11.89 & 16 32 39.28 & +78 11 53.6 & 33 & 162 & 1368 & 1.15 & 6.89 & 20.6 & -20.45 \\
NGC~7331  & Sbc & 10.20 & 22 37 04.16 & +34 24 56.0 & 75 & 169 & 818 & 4.89 & 20.06 & 14.1 & -21.68 \\
NGC~7678  & SBc & 12.50 & 23 28 27.87 & +22 25 14.0 & 44 &  21 & 3488 & 1.04 & 14.46 & 47.8 & -21.55 \\
\hline
\end{tabular}\\
\end{center}
\begin{flushleft}
$^a$ Coordinates of the galaxy center. Units of right ascension are hours, minutes, and
seconds, and units of declination are degrees, arcminutes, and arcseconds.
$^b$ Heliocentric radial velocity.
$^c$ Isophotal radius (25 mag/arcsec$^2$ in the $B$-band) corrected for Galactic 
extinction and absorption due to the inclination of a galaxy.
\end{flushleft}
\end{table*}

\begin{table}
\caption[]{\label{table:observ}
Journal of observations.
}
\begin{center}
\begin{tabular}{ccccc} \hline \hline
Object (Pos.$^a$) & Date & Exposures & Seeing & Airmass \\
 & & (s) & (arcsec) &  \\
\hline
NGC~628  (1) & 2008.02.07 & $900\times2$ & 1.9 & 2.03 \\
NGC~783  (1) & 2006.10.19 & $900\times3$ & 3.0 & 1.28 \\
NGC~783  (2) & 2007.09.07 & $900\times4$ & 1.6 & 1.02 \\
NGC~783  (3) & 2008.02.09 & $900\times6$ & 1.9 & 1.29 \\
NGC~2336 (1) & 2006.10.19 & $900\times4$ & 2.0 & 1.39 \\
NGC~2336 (2) & 2008.02.08 & $900\times8$ & 1.4 & 1.31 \\
NGC~2336 (3) & 2006.10.19 & $300+900$    & 2.0 & 1.32 \\
NGC~2336 (4) & 2008.02.09 & $900\times6$ & 1.2 & 1.33 \\
NGC~6217 (1) & 2006.08.23 & $900\times4$ & 1.8 & 1.47 \\
NGC~7331 (1) & 2006.08.23 & $900\times4$ & 2.1 & 1.17 \\
NGC~7678 (1) & 2006.08.22 & $900\times3$ & 1.4 & 1.15 \\
NGC~7678 (2) & 2006.10.19 & $900\times2$ & 1.6 & 1.08 \\
NGC~7678 (3) & 2006.10.19 & $900\times3$ & 1.6 & 1.08 \\
\hline
\end{tabular}\\
\end{center}
\begin{flushleft}
$^a$ Slit position \\
\end{flushleft}
\end{table}

Here we report spectra of 63 H\,{\sc ii} regions in a sample 
of six spiral galaxies:
NGC~628, NGC~783, NGC~2336, NGC~6217, NGC~7331, and NGC~7678.
A first motivation of the spectral observations is the 
determination of the oxygen and nitrogen abundances and
electron temperatures in individual  H\,{\sc ii} regions using the recent
NS- and ON calibration \citep{pilyugin2010,pilyuginmattsson2011}. 
The abundances will be used in the computation of the grid of 
evolutionary models of star clusters embedded in studied H\,{\sc ii} regions. 
The second motivation is the determination of the optical extinction
by dust to the gas, from the measured Balmer decrement in the
studied H\,{\sc ii} regions.

In this paper we describe observations and data reduction, 
the derivation of elemental abundances, electron temperatures and
optical extinction for individual H\,{\sc ii} regions and the 
examination their radial gradients across the disks.
Results of these examinations provide information about the chemical 
abundance of interstellar medium from which embedded stars formed, 
as well as the dust extinction estimations in the surrounding gas.

The paper is organized as follows.
The observations and data reduction are described in Section~2.
The oxygen and nitrogen abundances and the electron temperatures for
individual  H\,{\sc ii} regions as well 
the parameters of the radial distributions 
(the extrapolated central intercept value and the gradient) 
of the oxygen and nitrogen abundances and electron
temperature in galaxies are discussed in Section~3.
Section~4 gives a brief summary. 

\begin{figure}
\resizebox{1.00\hsize}{!}{\includegraphics[angle=000]{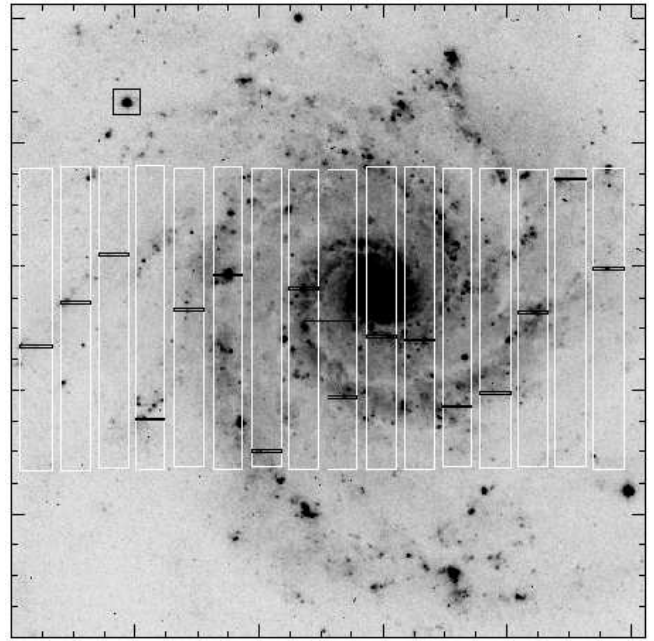}}
\caption{
Disposition of 16 slits (black small horizontal rectangles) for NGC~628. 
North is upward and east is to the left (P.A.$_{slits} = 90\degr$). 
The size of the image is $6.1\times6.1$ arcmin$^2$. The black square 
shows a guide star for the exposures.
}
\label{figure:fig1}
\end{figure}

Throughout the paper, we will be using the following notations for the line
fluxes,
\begin{equation}
R_2 = [{\rm O}\,\textsc{ii}] \lambda 3727+ \lambda 3729
    = I_{\rm [OII] \lambda 3727+ \lambda 3729} /I_{{\rm H}\beta },
\end{equation}
\begin{equation}
N_2 = [{\rm N}\,\textsc{ii}] \lambda 6548+ \lambda 6584
    = I_{\rm [NII] \lambda 6548+ \lambda 6584} /I_{{\rm H}\beta },
\label{equation:n2}
\end{equation}
\begin{equation}
S_2 = [{\rm S}\,\textsc{ii}] \lambda 6717+ \lambda 6731
    = I_{\rm [SII] \lambda 6717 + \lambda 6731} /I_{{\rm H}\beta },
\end{equation}
\begin{equation}
R_3 = [{\rm O}\,\textsc{iii}] \lambda 4959+ \lambda 5007
    = I_{{\rm [OIII]} \lambda 4959 + \lambda 5007} /I_{{\rm H}\beta }.
\label{equation:r3}
\end{equation}
The electron temperatures $t$ are given in units of 10$^4$K.

\section{OBSERVATIONS AND REDUCTION}

\begin{table}
\caption[]{\label{table:flux1}
Offsets, galactocentric distances, radial velocities 
of H\,{\sc ii} regions.
}
\begin{center}
\begin{tabular}{ccccccc} \hline \hline
H\,{\sc ii} & Pos.$^a$ & Ref.$^b$ & N--S$^c$ & E--W$^c$ & $r^d$ & $v^e$ \\ 
region       &      &     & (arcsec) & (arcsec) & (kpc) & (km/s) \\ 
\hline
\multicolumn{7}{c}{NGC~628}                                          \\
 1 & 1 & 94, 54  & +12.5 & +129.0 & 4.55 & 620  \\
 2 & 1 & 92      & -12.0 &  +87.7 & 3.11 & 665 \\ 
 3 & 1 & 77, 58  & -64.3 & +36.5 & 2.58 & 611 \\ 
 4 & 1 & 75      & -65.4 & +45.0 & 2.77  & 638 \\ 
 5 & 1 & 73, 11  & -28.8 & +25.3 & 1.34 & 654 \\ 
 6 & 1 & 25, 6   & +0.2 & -43.0 & 1.51 & 655 \\ 
 7 & 1 & 62, 81  & -91.8 & -68.1 & 4.01 & 628 \\ 
 8 & 1 & 58      & -72.0 & -132.6 & 5.31 & 614 \\ 
 9 & 1 & 35      & +19.7 & -153.1 & 5.42 & 716 \\ 
10 & 1 & 38, (5) & -6.4 & -177.9 & 6.25 & 711 \\ 
\multicolumn{7}{c}{NGC~783}                                          \\
1  & 1 & --- & -24.3 & +16.0 & 10.60 & 5181 \\
2  & 1 & --- & -27.5 & +23.7 & 12.83 & 5202 \\
3  & 1 & --- &  -4.8 & -19.7 &  8.33 & 5192 \\ 
4  & 1 & --- & +19.5 & -29.1 & 11.96 & 5235 \\ 
5  & 2 & --- & -19.5 & +22.1 & 10.17 & 5270 \\ 
6  & 2 & --- & -20.0 & +25.6 & 11.15 & 5103 \\ 
7  & 2 & --- & +21.6 & -28.3 & 12.20 & 5153 \\ 
8  & 3 & --- &  +5.1 & -22.9 &  8.42 & 5233 \\ 
\multicolumn{7}{c}{NGC~2336}                                         \\
 1a & 1 & --- & -113.8 & +33.5 & 20.79 & 2011 \\
 1b & 2 & --- & -113.8 & +33.5 & 20.79 & 2048 \\
  2 & 1 & 28 &    -89.0 & -40.0 & 16.89 & 2006 \\ 
  3 & 1 & 27a &   -71.8 &  +0.2 & 11.30 & 2040 \\ 
 4a & 1 & 26 &  -47.8 & -10.5 & 7.79 & 2009 \\ 
 4b & 4 & 26 &  -47.8 & -10.5 & 7.79 & 2052 \\ 
  5 & 1 & 22 &    -27.5 &  -9.8 & 4.86 & 2028 \\ 
  6 & 1 & 17 &    +17.5 & +85.4 & 23.02 & 2264 \\ 
 7a & 1 &  5 &  +95.9 & +13.6 & 15.19 & 2432 \\ 
 7b & 2 &  5 &  +95.9 & +13.6 & 15.19 & 2410 \\ 
  8 & 2 & 27d &    -82.5 &  -9.5 & 12.99 & 2013 \\ 
  9 & 2 & --- &   -62.6 &  +8.1 & 10.27 & 1917 \\ 
10a & 2 & 19 &   +3.4 & +43.8 & 11.82 & 2141 \\ 
10b & 4 & 19 &   +3.4 & +43.8 & 11.82 & 2198 \\ 
 11 & 2 & 16 &    +18.2 & +38.7 & 10.59 & 2349 \\ 
 12 & 2 & 10 &    +52.3 &  +0.2 & 8.22 & 2512 \\ 
 13 & 2 &  8 &    +68.1 & -34.5 & 14.84 & 2485 \\ 
14a & 2 &  2 & +142.0 & -11.5 & 22.83 & 2573 \\ 
14b & 4 &  2 & +142.0 & -11.5 & 22.83 & 2545 \\ 
15a & 3 & 30 & -109.0 & -32.1 & 18.49 & 1940 \\ 
15b & 4 & 30 & -109.0 & -32.1 & 18.49 & 1935 \\ 
16a & 3 & 27b &  -77.3 &  -3.6 & 12.11 & 2014 \\ 
16b & 4 & 27b &  -77.3 &  -3.6 & 12.11 & 2018 \\ 
17a & 3 & 21 &  -26.1 & +48.3 & 14.05 & 2167 \\ 
17b & 4 & 21 &  -26.1 & +48.3 & 14.05 & 2061 \\ 
 18 & 4 & 29 &    -94.5 & -37.3 & 17.18 & 2233 \\ 
 19 & 4 & 27e &    -83.9 &  -4.3 & 13.13 & 2064 \\ 
 20 & 4 & --- &    -31.6 & +48.6 & 14.49 & 2160 \\ 
 21 & 4 & --- &    -19.3 & +47.6 & 13.50 & 1982 \\ 
 22 & 4 & --- &    +27.2 & -62.7 & 17.88 & 2364 \\ 
 23 & 4 & 15 &    +24.4 & -61.7 & 17.47 & 2263 \\ 
 24 & 4 &  9 &    +62.6 & +48.3 & 15.61 & 2375 \\ 
 25 & 4 & --- &    +67.7 & +44.5 & 15.32 & 2443 \\ 
 26 & 4 & --- &    +77.7 &  -5.3 & 12.43 & 2575 \\ 
 27  & 4 &  7 &    +80.8 &  -4.6 & 12.88 & 2422 \\ 
 28  & 4 &  4 &  +111.7 & -31.1 & 20.17 & 2335 \\ 
\multicolumn{7}{c}{NGC~6217}                                         \\
1 & 1 & --- & +29.8 & +13.9 & 3.30 & 1443 \\ 
2 & 1 & --- & +42.2 & +13.0 & 4.41 & 1485 \\ 
3 & 1 & --- & -14.6 & +23.2 & 3.24 & 1348 \\ \hline
\end{tabular}\\
\end{center}
\end{table}

\setcounter{table}{2}
\begin{table}
\caption[]{Continued}
\begin{center}
\begin{tabular}{ccccccc} \hline \hline
H\,{\sc ii} & Pos.$^a$ & Ref.$^b$ & N--S$^c$ & E--W$^c$ & $r^d$ & $v^e$ \\ 
region       &      &     & (arcsec) & (arcsec) & (kpc) & (km/s) \\ 
\hline
\multicolumn{7}{c}{NGC~7331}                                         \\
1 & 1 & --- & +123.2 & +40.4 & 9.78 & 537 \\ 
2 & 1 & --- &  +76.8 & +18.0 & 5.45 & 512 \\ 
3 & 1 & --- &  +27.4 & +16.2 & 3.48 & 667 \\ 
4 & 1 &   1 &  -50.2 & +23.1 & 9.06 & 865 \\ 
\multicolumn{7}{c}{NGC~7678}                                         \\
 1 & 1 & --- & -54.2 & +12.6 & 13.02 & 3422 \\ 
 2 & 1 & --- & -31.8 &  -7.1 & 8.57 & 3431 \\ 
 3 & 1 & --- & -23.3 & -14.8 & 8.10 & 3525 \\ 
 4 & 1 & --- & -14.5 & -21.0 & 8.11 & 3553 \\ 
 5 & 1 & --- &  +3.7 & -26.6 & 8.15 & 3455 \\ 
 6 & 2 & --- & -26.5 & +14.2 & 7.02 & 3412 \\ 
 7 & 2 & --- & -10.5 & +31.6 & 9.63 & 3415 \\ 
8a & 2 & --- & -15.8 & +27.3 & 8.56 & 3426 \\ 
8b & 3 & --- & -15.8 & +27.3 & 8.56 & 3432 \\ 
 9 & 3 & --- & -46.5 &  -5.2 & 11.87 & 3384 \\ 
10 & 3 & --- & -23.0 &  +6.2 & 5.56  & 3398 \\ 
\hline
\end{tabular} 
\end{center}
\begin{flushleft}
$^a$ Slit position.  
$^b$ H\,{\sc ii} region number from \citet{belley1992} (first number), 
from \citet{rosales2011} (second number), and from \citet{vanzee1998} 
(second number in brackets) for NGC~628, from \citet{gusev2003} for 
NGC~2336, and from \citet{bresolin1999} for NGC~7331. 
$^c$ Offsets from the galactic centre (see Table~\ref{table:sample}), 
positive to the north and west. 
$^d$ Deprojected galactocentric distance. \\
$^e$ Radial velocity of H\,{\sc ii} region (km s$^{-1}$).
\end{flushleft}
\end{table}

We selected galaxies with known $UBVRI$ photometry of H\,{\sc ii} regions, 
obtained by our team in the previous works: NGC~628 \citep{bruevich2007}, 
NGC~783  \citep{gusev2006a,gusev2006b}, NGC~2336  \citep{gusev2003},
NGC~6217 \citep{artamonov1999}, NGC~7331 (unpublished), 
NGC~7678 \citep{artamonov1997}. 
The galaxy sample is presented in Table~\ref{table:sample}. 
The columns show parameters from the {\sc leda} data base 
\citep{paturel2003}: morphological type and apparent magnitude of the galaxy 
in columns (2) and (3), coordinates (epoch 2000) in columns (4) and (5),
the inclination and position angles in columns (6) and (7), radial velocity 
in column (8), the isophotal radius in arcmin and kpc in columns (9) and (10),
distance  in the column (11), the absolute blue magnitude $M_B$ in column (12).

\subsection{Observations}

The observations were carried out at the 6 m telescope of Special Astrophysical 
Obsevatory (SAO) of the Russian Academy of Sciences
with Spectral Camera attached at the focal reducer SCORPIO 
\citep{afanasiev2005} ($f/4 \to f/2.6$) in the multislit mode; 
the field was about 6 arcmin and the pixel size was 0.178 arcsec on a 
EEV~42-40 ($2048\times2048$ pixels) CCD detector.
The data for six galaxies were acquired during observing runs in 2006--2008
(see the journal of observations in Table~\ref{table:observ}). 

The SCORPIO multislit unit is an arrangement that consists of 16 metal 
strips with slits located in the focal plane and moved in a 
$2.9\times5.9$ arcmin$^2$ field (Fig.~\ref{figure:fig1}). 
The size of slits is $1.5\times18$ arcsec$^2$, the distance between centres of
neighbouring slits is 22 arcsec.

\begin{figure}
\vspace{0.4cm}
\resizebox{1.00\hsize}{!}{\includegraphics[angle=000]{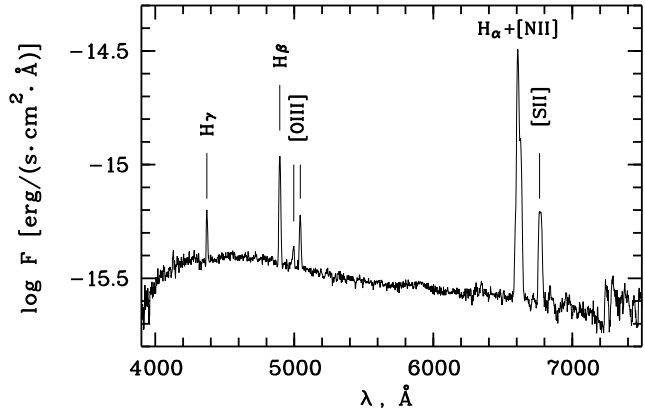}}
\caption{
The observed spectrum ''b'' of the H\,{\sc ii} region 17 in NGC~2336.
}
\label{figure:fig2}
\end{figure}

We used the grism VPHG550G with a dispersion of 2.1\AA/pixel and
a spectral resolution of 10\AA, which provided spectral coverage 
from [O\,{\sc ii}]$\lambda$3727+$\lambda$3729 oxygen emission lines 
to [S\,{\sc ii}]$\lambda$6717+$\lambda$6731 sulphur emission lines. The 
example of spectrum is presented in Fig.~\ref{figure:fig2}.

We selected target H\,{\sc ii} regions using $B$ band and H$\alpha$ CCD images
of galaxies, which were obtained with the 1 m and 1.5 m telescopes of 
the Mt. Maidanak Observatory (Institute of Astronomy, Uzbek Academy of 
Sciences) in Uzbekistan and with the 1.8 m telescope of the Bohyunsan Optical
Astronomy Observatory (Korea Astronomy and Space Science Institute) 
in Korea. 

As mentioned above, the broadband $UBVRI$ photometry
of selected H\,{\sc ii} regions, derived previously by us and
our colleagues as a part of our study of star formation processes
in spiral and irregular galaxies, provided the source of targets
for the multi-slit observations. We selected objects widely
distributed across the face of the galaxies, enabling the study of
the chemical abundance gradients. The H\,{\sc ii} regions were selected
for spectroscopy based on their brightness and size, with
typical angular diameters from 2 up to 5 arcsec.

The total exposure time of the set consists of several time intervals. 
After the every time interval, positions of slits were displaced along the X axes 
with a step of 20 pixels. X axes runs along the slit and Y axes runs 
perpendicularly to the slit (see Fig.~\ref{figure:fig1}).
This technique allows us to obtain spectra of several neighbouring 
H\,{\sc ii} regions using the same slit. 
Therefore, the exposure time for individual H\,{\sc ii} regions was smaller 
than the total exposure time of the set indicated in the 
Table~\ref{table:observ}.

The observing procedure consisted of obtaining bias, flat and wavelength 
calibration images at the beginning and the end of the each set.
Several spectrophotometric standards were observed during the each 
night at different air mass.

\begin{table*}
\caption[]{\label{table:flux2} The reddening-corrected fluxes of main emission
lines of H\,{\sc ii} regions.}
\begin{center}
\begin{tabular}{ccccccc} \hline \hline
H\,{\sc ii} & [O\,{\sc ii}]$^a$ & [O\,{\sc iii}]$^a$ & [N\,{\sc ii}]$^a$ & [S\,{\sc ii}]$^a$ & $F$(H$\beta$)$^b$ & $c$(H$\beta$) \\
region & 3727+3729 & 5007 & 6584 & 6717+6731 & 4861 &  \\
\hline
\multicolumn{7}{c}{NGC~628} \\
 1  &      ---      &      ---      & 0.73$\pm$0.41 & 0.42$\pm$0.24 &   7.48$\pm$1.17 & 0.74$\pm$0.31 \\
 2  &      ---      &      ---      & 1.06$\pm$0.42 & 1.16$\pm$0.45 &   8.04$\pm$0.85 & 0.58$\pm$0.24 \\
 3  &      ---      &      ---      & 0.61$\pm$0.26 & 0.35$\pm$0.14 &  27.40$\pm$3.07 & 0.33$\pm$0.21 \\
 4  &      ---      & 0.31$\pm$0.13 & 0.46$\pm$0.35 & 0.54$\pm$0.30 &   5.21$\pm$0.79 & 0.37$\pm$0.31 \\
 5  &      ---      &      ---      & 0.70$\pm$0.35 & 0.78$\pm$0.35 &   9.61$\pm$1.08 & 0.30$\pm$0.25 \\
 6  &      ---      &      ---      &      ---      &      ---      &   2.12$\pm$0.80 & 0.85$\pm$0.96 \\
 7  & 5.64$\pm$2.01 & 0.49$\pm$0.07 & 0.79$\pm$0.22 & 0.47$\pm$0.12 &  33.32$\pm$1.82 & 0.00$\pm$0.13 \\
 8  & 4.06$\pm$1.74 & 0.66$\pm$0.07 & 0.77$\pm$0.18 & 0.51$\pm$0.10 &  38.15$\pm$2.05 & 0.12$\pm$0.12 \\
 9  &      ---      & 1.62$\pm$0.18 & 0.59$\pm$0.16 & 0.32$\pm$0.08 &  22.62$\pm$1.41 & 0.87$\pm$0.13 \\
10  &      ---      & 0.36$\pm$0.10 & 0.74$\pm$0.29 & 0.77$\pm$0.30 &  30.51$\pm$3.80 & 0.03$\pm$0.22 \\
\multicolumn{7}{c}{NGC~783} \\
 1  &      ---      & 1.47$\pm$0.20 & 1.06$\pm$0.27 & 0.72$\pm$0.20 &   5.81$\pm$0.38 & 0.28$\pm$0.15 \\
 2  &      ---      & 0.73$\pm$0.30 &      ---      &      ---      &   1.38$\pm$0.36 & 0.08$\pm$0.55 \\
 3  & 1.80$\pm$0.60 & 0.23$\pm$0.04 & 1.33$\pm$0.20 & 0.89$\pm$0.13 &  22.15$\pm$6.51 & 0.51$\pm$0.09 \\
 4  &      ---      & 0.68$\pm$0.13 & 1.46$\pm$0.46 & 1.19$\pm$0.39 &   3.34$\pm$0.29 & 0.90$\pm$0.20 \\
 5  & 2.34$\pm$0.68 & 0.23$\pm$0.04 & 0.54$\pm$0.14 & 0.86$\pm$0.15 &   7.78$\pm$0.31 & 0.61$\pm$0.10 \\
 6  & 2.45$\pm$0.53 & 0.44$\pm$0.05 & 1.50$\pm$0.21 & 1.16$\pm$0.17 &  11.69$\pm$0.42 & 0.44$\pm$0.09 \\
 7  &      ---      & 1.25$\pm$0.46 & 3.11$\pm$1.87 & 2.30$\pm$1.46 &   2.55$\pm$0.46 & 0.76$\pm$0.45 \\
 8  & 2.35$\pm$0.54 & 0.32$\pm$0.06 & 1.16$\pm$0.19 & 0.64$\pm$0.12 &  25.97$\pm$1.00 & 0.73$\pm$0.09 \\
\multicolumn{7}{c}{NGC~2336} \\
 1a &      ---      & 0.61$\pm$0.23 & 1.15$\pm$0.59 & 1.05$\pm$0.53 &   2.04$\pm$0.35 & 0.46$\pm$0.33 \\
 1b & 4.16$\pm$1.07 & 1.01$\pm$0.11 & 0.89$\pm$0.18 & 1.37$\pm$0.24 &   6.80$\pm$0.33 & 0.15$\pm$0.11 \\
 2  & 2.26$\pm$0.66 & 0.58$\pm$0.05 & 0.81$\pm$0.12 & 0.51$\pm$0.07 &  22.07$\pm$0.77 & 0.45$\pm$0.08 \\
 3  &      ---      & 0.27$\pm$0.07 & 0.80$\pm$0.23 & 0.44$\pm$0.14 &  16.36$\pm$1.18 & 0.69$\pm$0.16 \\
 4a &      ---      &      ---      & 0.66$\pm$0.16 & 0.47$\pm$0.10 &   7.83$\pm$0.40 & 0.71$\pm$0.12 \\
 4b &      ---      &      ---      & 0.82$\pm$0.22 & 0.57$\pm$0.15 &  22.58$\pm$1.29 & 0.34$\pm$0.14 \\
 5  &      ---      &      ---      & 0.91$\pm$0.56 &      ---      &   2.70$\pm$0.31 & 0.34$\pm$0.32 \\
 6  &      ---      & 1.42$\pm$0.06 & 0.48$\pm$0.08 & 0.62$\pm$0.07 &  19.92$\pm$0.53 & 1.12$\pm$0.06 \\
 7a &      ---      & 0.48$\pm$0.09 & 0.85$\pm$0.23 & 0.66$\pm$0.19 &   5.37$\pm$0.38 & 0.46$\pm$0.15 \\
 7b &      ---      & 0.27$\pm$0.02 & 0.88$\pm$0.11 & 0.66$\pm$0.08 &  35.97$\pm$0.98 & 0.45$\pm$0.06 \\
 8  & 2.21$\pm$0.48 & 0.14$\pm$0.02 & 0.71$\pm$0.11 & 0.66$\pm$0.08 &  28.12$\pm$0.70 & 0.32$\pm$0.07 \\
 9  &      ---      & 0.05$\pm$0.01 & 0.77$\pm$0.13 & 0.60$\pm$0.09 &  20.06$\pm$0.56 & 0.11$\pm$0.08 \\
10a &      ---      & 0.16$\pm$0.03 & 1.07$\pm$0.17 & 0.80$\pm$0.13 &  18.14$\pm$0.63 & 0.46$\pm$0.09 \\
10b &      ---      & 0.17$\pm$0.02 & 1.01$\pm$0.13 & 0.70$\pm$0.08 &  33.81$\pm$0.96 & 0.64$\pm$0.07 \\
11  &      ---      & 0.47$\pm$0.08 & 1.17$\pm$0.27 & 1.14$\pm$0.26 &   8.42$\pm$0.55 & 0.90$\pm$0.14 \\
12  &      ---      &      ---      & 0.85$\pm$0.36 & 0.79$\pm$0.34 &   4.94$\pm$0.62 & 0.93$\pm$0.26 \\
13  &      ---      & 0.35$\pm$0.04 & 1.04$\pm$0.15 & 0.60$\pm$0.09 &  11.16$\pm$0.38 & 1.23$\pm$0.08 \\
14a &      ---      & 1.15$\pm$0.10 & 0.91$\pm$0.16 & 0.98$\pm$0.16 &  18.31$\pm$0.82 & 0.72$\pm$0.10 \\
14b & 3.36$\pm$0.76 & 1.18$\pm$0.07 & 0.79$\pm$0.11 & 0.87$\pm$0.10 &  20.25$\pm$0.66 & 0.70$\pm$0.07 \\
15a &      ---      & 1.68$\pm$0.54 & 0.65$\pm$0.51 &      ---      &   2.47$\pm$0.44 & 0.21$\pm$0.40 \\
15b & 6.36$\pm$2.07 & 0.89$\pm$0.11 & 0.66$\pm$0.18 & 0.78$\pm$0.20 &   9.32$\pm$0.47 & 0.07$\pm$0.13 \\
16a & 2.07$\pm$0.92 & 0.32$\pm$0.08 & 0.73$\pm$0.20 & 0.79$\pm$0.20 &   9.31$\pm$0.65 & 0.66$\pm$0.15 \\
16b & 2.63$\pm$0.61 & 0.21$\pm$0.03 & 0.73$\pm$0.12 & 0.68$\pm$0.09 &  26.12$\pm$0.75 & 0.56$\pm$0.07 \\
17a &      ---      & 0.25$\pm$0.06 & 0.92$\pm$0.25 & 0.73$\pm$0.20 &  15.98$\pm$1.32 & 0.38$\pm$0.16 \\
17b &      ---      & 0.32$\pm$0.02 & 1.22$\pm$0.11 & 0.65$\pm$0.05 &  99.43$\pm$1.82 & 0.62$\pm$0.05 \\
18  &      ---      &      ---      & 0.75$\pm$0.47 & 0.89$\pm$0.56 &   3.63$\pm$0.66 & 1.32$\pm$0.38 \\
19  & 7.54$\pm$2.31 & 0.23$\pm$0.07 & 0.72$\pm$0.20 & 0.82$\pm$0.20 &  13.78$\pm$0.88 & 0.71$\pm$0.14 \\
20  &      ---      & 0.24$\pm$0.08 & 1.11$\pm$0.25 & 0.89$\pm$0.21 &   8.53$\pm$0.57 & 1.06$\pm$0.13 \\
21  &      ---      & 0.30$\pm$0.05 & 1.73$\pm$0.31 & 0.94$\pm$0.20 &   8.93$\pm$0.43 & 0.35$\pm$0.12 \\
22  & 1.78$\pm$0.29 & 0.53$\pm$0.02 & 0.81$\pm$0.08 & 0.60$\pm$0.05 &  22.86$\pm$0.44 & 0.49$\pm$0.05 \\
23  & 1.88$\pm$0.25 & 0.68$\pm$0.02 & 0.85$\pm$0.08 & 0.60$\pm$0.04 &  42.42$\pm$0.74 & 0.09$\pm$0.04 \\
24  &      ---      & 0.76$\pm$0.20 & 1.33$\pm$0.52 & 1.24$\pm$0.52 &   3.62$\pm$0.42 & 0.84$\pm$0.25 \\
25  &      ---      & 0.82$\pm$0.16 & 0.87$\pm$0.25 & 0.71$\pm$0.22 &   5.12$\pm$0.41 & 0.86$\pm$0.17 \\
26  &      ---      & 0.47$\pm$0.22 & 1.34$\pm$0.64 & 1.13$\pm$0.55 &   5.96$\pm$0.93 & 0.92$\pm$0.32 \\
27  &      ---      & 0.72$\pm$0.22 & 1.39$\pm$0.56 & 1.21$\pm$0.54 &   4.56$\pm$0.44 & 0.50$\pm$0.25 \\
28  & 2.37$\pm$0.40 & 0.70$\pm$0.03 & 0.89$\pm$0.09 & 0.69$\pm$0.08 &  38.60$\pm$0.81 & 0.08$\pm$0.05 \\
\multicolumn{7}{c}{NGC~6217} \\
 1  & 1.87$\pm$0.81 & 0.25$\pm$0.05 & 0.85$\pm$0.16 & 0.52$\pm$0.10 &  21.24$\pm$0.86 & 0.10$\pm$0.10 \\
 2  &      ---      &      ---      & 1.09$\pm$0.52 & 0.89$\pm$0.42 &   7.37$\pm$1.14 & 0.76$\pm$0.31 \\
\hline
\end{tabular}
\end{center}
\begin{flushleft}
$^a$ $I$($\lambda$)/$I$(H$\beta$) ratio.
$^b$ The fluxes are in units of $10^{-16}$ erg \,s$^{-1}$\,cm$^{-2}$. \\
\end{flushleft}
\end{table*}

\setcounter{table}{3}
\begin{table*}
\caption[]{Continued
}
\begin{center}
\begin{tabular}{ccccccc} \hline \hline
H\,{\sc ii} & [O\,{\sc ii}]$^a$ & [O\,{\sc iii}]$^a$ & [N\,{\sc ii}]$^a$ & [S\,{\sc ii}]$^a$ & $F$(H$\beta$)$^b$ & $c$(H$\beta$) \\
region & 3727+3729 & 5007 & 6584 & 6717+6731 & 4861 & \\
\hline
\multicolumn{7}{c}{NGC~6217} \\
 3  &      ---      & 0.98$\pm$0.31 & 1.08$\pm$0.43 & 0.71$\pm$0.29 &   7.70$\pm$1.03 & 0.80$\pm$0.25 \\
\multicolumn{7}{c}{NGC~7331} \\
 1  &      ---      & 1.05$\pm$0.22 & 1.20$\pm$0.57 & 0.91$\pm$0.44 &   3.16$\pm$0.36 & 0.09$\pm$0.28 \\
 2  &      ---      &      ---      & 2.65$\pm$2.33 & 0.91$\pm$0.90 &   1.13$\pm$0.31 & 1.53$\pm$0.68 \\
 3  &      ---      & 0.14$\pm$0.06 & 0.44$\pm$0.16 & 0.26$\pm$0.09 &   9.27$\pm$0.79 & 1.24$\pm$0.17 \\
 4  & 2.32$\pm$1.15 & 0.41$\pm$0.04 & 0.99$\pm$0.14 & 0.82$\pm$0.10 &  26.64$\pm$0.88 & 0.50$\pm$0.08 \\
\multicolumn{7}{c}{NGC~7678} \\
 1  &      ---      & 1.17$\pm$0.58 & 0.82$\pm$0.68 & 0.88$\pm$0.71 &   3.03$\pm$0.81 & 0.94$\pm$0.51 \\
 2  &      ---      & 0.52$\pm$0.14 & 1.01$\pm$0.36 & 0.83$\pm$0.29 &   6.34$\pm$0.59 & 0.32$\pm$0.21 \\
 3  & 2.74$\pm$0.91 & 0.30$\pm$0.05 & 0.48$\pm$0.13 & 0.67$\pm$0.14 &  28.74$\pm$1.30 & 0.14$\pm$0.10 \\
 4  & 2.00$\pm$0.54 & 0.49$\pm$0.05 & 0.74$\pm$0.13 & 0.60$\pm$0.10 &  29.05$\pm$1.10 & 0.30$\pm$0.08 \\
 5  & 4.07$\pm$1.78 & 0.39$\pm$0.12 & 1.47$\pm$0.42 & 0.77$\pm$0.27 &   7.57$\pm$0.60 & 0.06$\pm$0.19 \\
 6  &      ---      & 0.31$\pm$0.08 & 1.05$\pm$0.24 & 0.97$\pm$0.21 &   9.54$\pm$0.61 & 0.77$\pm$0.14 \\
 7  &      ---      & 0.63$\pm$0.03 & 0.89$\pm$0.09 & 0.85$\pm$0.06 &  61.66$\pm$1.21 & 0.69$\pm$0.05 \\
 8a &      ---      & 0.54$\pm$0.02 & 0.85$\pm$0.08 & 0.69$\pm$0.04 & 196.70$\pm$3.16 & 0.80$\pm$0.04 \\
 8b &      ---      & 0.50$\pm$0.02 & 0.91$\pm$0.08 & 0.62$\pm$0.04 & 101.40$\pm$1.63 & 0.90$\pm$0.04 \\
 9  &      ---      & 0.67$\pm$0.04 & 1.04$\pm$0.13 & 0.63$\pm$0.08 &  17.35$\pm$0.52 & 0.79$\pm$0.07 \\
10  & 2.53$\pm$0.44 & 0.44$\pm$0.02 & 0.95$\pm$0.10 & 0.93$\pm$0.08 &  41.05$\pm$0.81 & 0.71$\pm$0.05 \\
\hline
\end{tabular}
\end{center}
\begin{flushleft}
$^a$ $I$($\lambda$)/$I$(H$\beta$) ratio.
$^b$ The fluxes are in units of $10^{-16}$ erg \,s$^{-1}$\,cm$^{-2}$. \\
\end{flushleft}
\end{table*}

\subsection{Data reduction}

Initial data reduction followed routine procedures, including bias, 
cosmic-ray,
flat-field and atmospheric extinction corrections and photometric calibration, 
intensity normalisation to the intensity of the central ($8^{th}$) slit, 
wavelength calibration with a standard He-Ne-Ar lamp,
subtracting the background, transformation to one-dimensional
spectrum and summing of spectra, 
using the European Southern Observatory Munich Image Data Analysis
System (MIDAS).

The emission line fluxes were measured using the 
continuum-substracted spectrum. 
Flux calibration was performed, using standard stars
BD+25$\degr$4655, HZ2, HZ4, Feige~34, and G193--74 \citep{oke1990}.
Both, the spectra of standard stars and the SAO astroclimate data of
\cite{kartasheva1978} were used for a calculation of the extinction
coefficient and a correction for the atmospheric extinction.
The blended lines H$\alpha$+[N\,{\sc ii}]$\lambda$6548+$\lambda$6584 and 
[S\,{\sc ii}]$\lambda$6717+$\lambda$6731 were measured by the three or 
two-Gaussian fitting, respectively. 

The extraction aperture corresponded to the area, where 
brightest emission lines from H\,{\sc ii} regions were 
''visible'' above the noise. This size is close to 
the angular diameter of individual H\,{\sc ii} regions from the 
imaging surveys as projected along the P.A. of the slit.

Coordinates, deprojected galactocentric distances, and radial velocities
are listed in Table~\ref{table:flux1}. 
Reddening-corrected line intensities $I$($\lambda$)/$I$(H$\beta$), 
undereddened fluxes $F$(H$\beta$), and the logarithmic extinction 
coefficients $c$(H$\beta$) are given in Table~\ref{table:flux2}.
Though the sulphur lines [S\,{\sc ii}]$\lambda$6717 and [S\,{\sc ii}]$\lambda$6731 
were deblended in several cases, the accuracy was very low. Therefore the
Table~\ref{table:flux2} presents the sum of the
[S\,{\sc ii}]($\lambda$6717+$\lambda$6731) of sulphur lines.
Some H\,{\sc ii} regions were observed twice, they are marked with letters 
''a'' and ''b'' in Tables~\ref{table:flux1},~\ref{table:flux2}. 

Absolute emission-line fluxes derived for the same object in different nights 
can be differ because of following reasons: a) different seeings in different 
nights and b) a slight deviation of the slit from the previous position. 
Note that the apparent sizes of the studied H\,{\sc ii} regions are
generally larger than the width of the slit. It should
be noted that, although the absolute emission-line fluxes can differ
by a large amount between different observations depending of
the exact slit placement, flux ratios derived for the same object on
different nights coincide within the errors (see Table~\ref{table:flux2}).

At calculation of the error of the line intensity measurement the 
following factors have been taken into consideration. The first 
factor is related to the Poisson statistics of the line photon flux. 
The second factor of the error appears by the computation of the 
underlying continuum, and makes the main contribution to the unaccuracy 
of faint lines. The third factor concerns with the uncertainty of the 
spectral sensitivity curve. It gives an additional error to the relative 
line intensities. The last factor is related to the goodness of fit of 
the line profile and is crucial for blended lines. All these components 
are summed in quadrature. The total errors have been propagated to 
calculate the errors of all derived parameters.

Estimations of radial velocities of studied H\,{\sc ii} regions (last column 
in Table ~\ref{table:flux1}) were derived as a by-product through the 
measurements of Doppler shifts of H$\alpha$ line in spectra. The accuracy of 
radial velocity measurements is about 30 km/s. Analysis of the observed radial 
velocities is beyond of the scope of this paper.

The measured emission fluxes $F$ were corrected for the interstellar 
reddening and Balmer absorption in the underlying stellar continuum. 
We used the theoretical H$\alpha$ to H$\beta$ ratio 
from \cite{osterbrock1989} assuming case B recombination and 
an electron temperature of 10,000~K and the analytical approximation 
to the Whitford interstellar reddening law by \citet{izotov1994}. 
We adopted the absorption equivalent width EW$_a$($\lambda$) = 2\AA \,
for hydrogen lines 
for all objects, which is the mean value for H\,{\sc ii} regions
derived by \cite{mccall1985}. For lines other than hydrogen
EW$_a$($\lambda$) = 0. 
The uncertainty of the logarithmic extinction coefficient $c$(H$\beta$)
is calculated from the measurements errors of H$\alpha$ and H$\beta$
lines and propagated to the dereddened line ratios.

\begin{figure}
\vspace{0.3cm}
\resizebox{1.00\hsize}{!}{\includegraphics[angle=000]{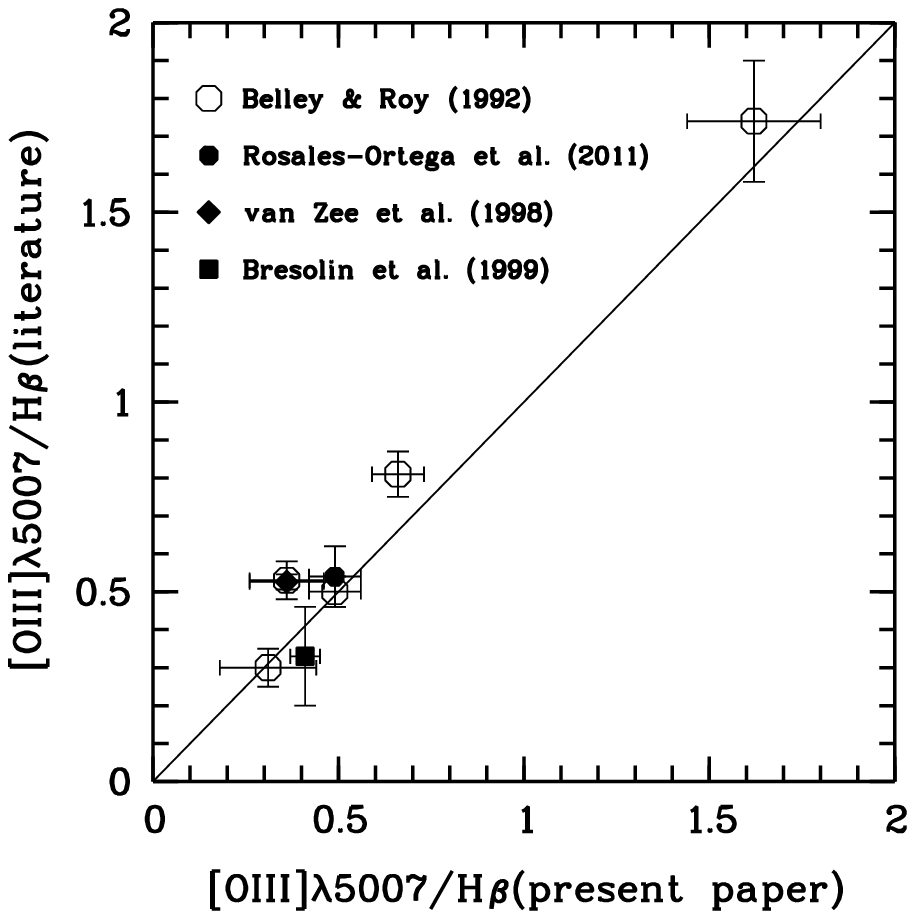}}
\caption{
Comparison between redding-corrected fluxes 
[O\,{\sc iii}]$\lambda$5007/H$\beta$ obtained in the present paper and 
data from the literature \citep{belley1992,rosales2011,vanzee1998,bresolin1999} 
for common H\,{\sc ii} regions.
}
\label{figure:liter}
\end{figure}

\subsection{Comparison with previous observations}

H\,{\sc ii} regions in the galaxies NGC~628 and NGC~7331 have previously
been observed via spectroscopy
\citep{mccall1985,ferguson1998,vanzee1998,bresolin1999} or the 
spectrophotometric imaging \citep{belley1992} or the 
Integral Field Spectroscopy method \citep{rosales2011}.
There is only one overlapping with the previous spectroscopic observations in every galaxy: 
one object in NGC~628 observed by \citet{vanzee1998} and one object in NGC~7331 
observed by \citet{bresolin1999}. 
In Fig.~\ref{figure:liter} we compared our measurements 
with results of these two previous spectroscopic observations, with results of the spectrophotometric 
imaging \citep{belley1992}, which are not as accurate as
spectroscopic ones for individual objects, and with results of the Integral Field
Spectroscopy for NGC~628 (one object \citep{rosales2011})
(see Table~\ref{table:flux1}).

Fig.~\ref{figure:liter} shows a satisfactory agreement between our spectroscopy and previous
observations. Note that for the quantitative analysis more statistics is
necessary. Previous spectroscopic observations of H\,{\sc ii} regions in
 NGC~628 and NGC~7331 will be also compared in Section~3 via radial
oxygen and nitrogen abundances distributions in disks of parent galaxies.

\section{Abundances}

Here we use the spectral data, presented in Section 2 for the determination of the 
oxygen and nitrogen abundances (and electron temperatures) in H\,{\sc ii} regions 
aiming to establish the radial distributions of these elements across the 
disks of galaxies.

\subsection{Preliminary remarks}

Since our measurements of the [N\,{\sc ii}]$\lambda$6584 and [O\,{\sc
iii}]$\lambda$5007 lines are more reliable than those of the
[N\,{\sc ii}]$\lambda$6548 and [O\,{\sc iii}]$\lambda$4959 lines,
we used [N\,{\sc ii}]$\lambda$6584 and [O\,{\sc iii}]$\lambda$5007 lines only. 
The [O\,{\sc iii}]$\lambda$5007 and $\lambda$4959 lines originate from transitions from the 
same energy level, so their fluxes ratio is due only to the transition probability ratio 
which is 3.013 \citep{storey2000}. Hence, the value of $R_3$ can be well approximated by    
\begin{equation}
R_3     = 1.33 \; I_{\rm {\rm [OIII]} \lambda 5007} /I_{\rm {\rm H}\beta }.
\label{equation:r3alternat}
\end{equation}
Similarly, the [N\,{\sc ii}]$\lambda$6584 and $\lambda$6548 lines also originate from transitions from the 
same energy level and the transition probability ratio for those lines is 3.071 
\citep{storey2000}. The value of $N_2$ is therefore well approximated by
\begin{equation}
N_2  = 1.33 \; I_{\rm \rm [NII] \lambda 6584} /I_{\rm {\rm H}\beta }.
\label{equation:n2alternat}
\end{equation}
We have used  Eq.(\ref{equation:n2alternat}) instead of Eq.(\ref{equation:n2})  to obtain the value of  
$N_2$ and  Eq.(\ref{equation:r3alternat}) instead of Eq.(\ref{equation:r3})  to determine  the value of $R_3$.

The intensities of strong, easily measured lines can be used to separate
different types of emission-line objects according to their main
excitation mechanism. \citet{baldwin1981}  proposed a diagram
(BPT classification diagram) where
the excitation properties of H\,{\sc ii} regions are studied by
plotting the low-excitation [N\,{\sc ii}]$\lambda$6584/H$\alpha$
line ratio against the high-excitation [O\,{\sc iii}]$\lambda$5007/H$\beta$
line ratio.

\begin{figure}
\resizebox{1.00\hsize}{!}{\includegraphics[angle=000]{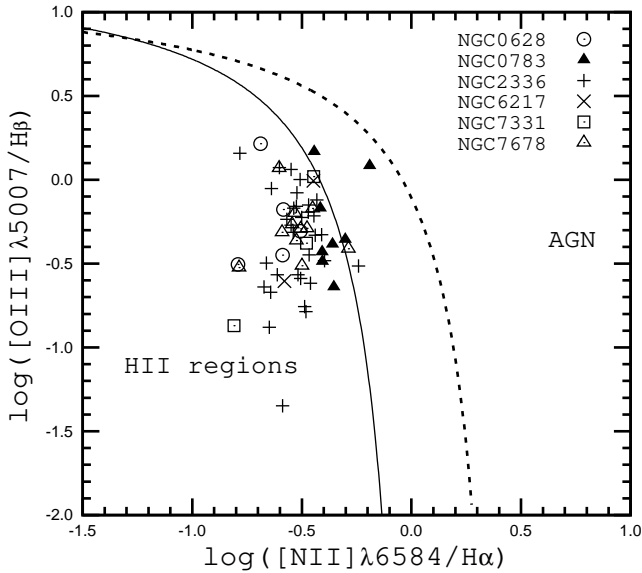}}
\caption{
The [N\,{\sc ii}]$\lambda$6584/H$\alpha$ versus [O\,{\sc iii}]$\lambda$5007/H$\beta$ diagram.
The symbols are the H\,{\sc ii} regions.
The solid line separates objects with H\,{\sc ii} spectra from those
containing an AGN according to \citet{kauffmann2003}. 
The dashed line is the separation line from \citet{kewleyetal2001}.
}
\label{figure:seagull}
\end{figure}

The [N\,{\sc ii}]$\lambda$6584/H$\alpha$ versus [O\,{\sc iii}]$\lambda$5007/H$\beta$ diagram
is shown in Fig.~\ref{figure:seagull}.
The symbols are  H\,{\sc ii} regions.
The solid line represents the relation
\begin{equation}
\log (\mbox{\rm [O\,{\sc iii}]$\lambda$5007/H$\beta$}) =
\frac{0.61}{\log (\mbox{\rm [[N\,{\sc ii}]$\lambda$6584/H$\alpha$})-0.05} +1.3,
\label{equation:kauff}
\end{equation}
which separates objects with H\,{\sc ii} spectra from those containing an AGN \citep{kauffmann2003}. 
The dashed separation line is the relation
\begin{equation}
\log (\mbox{\rm [O\,{\sc iii}]$\lambda$5007/H$\beta$}) =
\frac{0.61}{\log (\mbox{\rm [[N\,{\sc ii}]$\lambda$6584/H$\alpha$})-0.47} +1.19 
\label{equation:kewley}   
\end{equation}
from \citet{kewleyetal2001}.
Fig.~\ref{figure:seagull} shows that all H\,{\sc ii} regions from our sample are thermally 
photoionised objects if the separation line from \citet{kewleyetal2001} is used.
When the separation line from \citet{kauffmann2003} is used then   
one object (No.~7 in NGC~783) has an appreciable shift from the separation line 
towards the AGNs. However the uncertainty in the 
measurement of the intensity of the [N\,{\sc ii}]$\lambda$6584 line is large for this object 
(see Table~\ref{table:flux2}). 
Therefore all H\,{\sc ii} regions are included in further consideration. 
The abundances determined for object No.~7 from NGC~783 are not
considered in question.

\subsection{Abundances and temperatures}

Accurate oxygen and nitrogen abundances in H\,{\sc ii} regions can be derived 
via the classic $T_{e}$ method, often referred to as the direct method. 
This method is based on the measurement of the electron temperature $t_3$ 
within the [O\,{\sc iii}] zone and/or the electron temperature $t_2$ within the 
[O\,{\sc ii}] zone. The ratio of the nebular to auroral oxygen line intensities 
[O\,{\sc iii}]($\lambda$4959+$\lambda$5007)/[O\,{\sc iii}]$\lambda 4363$ 
is usually used for the $t_{3}$ determination, while 
the ratios of the nebular to auroral nitrogen line intensities 
[N\,{\sc ii}]($\lambda 6548+\lambda 6584$)/[N\,{\sc ii}]$\lambda 5755$ or 
[O\,{\sc ii}]($\lambda 3727+\lambda 3729$)/[O\,{\sc ii}]($\lambda 7320+\lambda 7330$) 
are used for the $t_{2}$ determination.
The auroral lines in spectra of H\,{\sc ii} regions from our sample are too faint  to be detected. 
In this case, it is still possible to obtain the estimates of the nebular abundances, using intensity ratios 
of the brightest ionic emission lines.

\citet{pagel1979} and \citet{alloinetal1979} have suggested that the 
locations of H\,{\sc ii} regions in some emission-line diagrams can be calibrated in 
terms of their oxygen abundances. This approach to abundance determination 
in H\,{\sc ii} regions, usually referred to as the ''strong line method'' has 
been widely adopted, especially in the  cases where the temperature-sensitive 
auroral lines are not detected. Numerous relations have been 
proposed to convert metallicity-sensitive emission-line ratios into metallicity or temperature estimates
\citep[e.g.][]{dopitaevans1986,zaritsky1994,vilchezesteban1996,pilyugin2000,pilyugin2001,
pettinipagel2004,tremonti2004,pilyuginthuan2005,liangetal2006,stasinska2006,thuan2010}. 
(See the review in \citet{ellison2008,lopezsanchez2010} for more details).

Two calibration relations for estimation of the oxygen and nitrogen abundances  
H\,{\sc ii} regions have been recently suggested. The ON-calibration relations 
give the oxygen and nitrogen abundances and electron temperature  in terms of 
the fluxes of the strong emission lines O$^{++}$, O$^{+}$, and N$^+$ \citep{pilyugin2010}. 
The NS-calibration relations give abundances and electron temperatures in terms 
of the fluxes in the strong emission lines of O$^{++}$, N$^+$, and S$^+$ \citep{pilyuginmattsson2011}. 
The ON and NS calibrations provide reliable oxygen and nitrogen
abundances for H\,{\sc ii} regions of all metallicities.
The oxygen $R_2$ line is measured in presented spectra with a large uncertainty 
(or even not available). Therefore we use the NS calibration method to estimate 
the abundances and electron temperatures in our sample of H\,{\sc ii} regions.

\begin{table}
\caption[]{\label{table:abun}
Oxygen and nitrogen abundances and electron temperatures $t_2$ in  H\,{\sc ii} regions derived
using the NS calibrations.
}
\begin{center}
\begin{tabular}{ccccc} \hline \hline
 H\,{\sc ii} region                                 &
 R$_G$$^a$                                              &
12+log(O/H)$_{\rm NS}$                                  &
12+log(N/H)$_{\rm NS}$                                  &
t$_{\rm NS}$$^b$                                      \\ 
\hline
\multicolumn{5}{c}{NGC~628}                      \\
    4  &   0.25 &    8.36 &    7.16 &    0.77 \\ 
    7  &   0.37 &    8.58 &    7.79 &    0.74 \\ 
    8  &   0.48 &    8.55 &    7.70 &    0.77 \\ 
    9  &   0.49 &    8.56 &    7.57 &    0.82 \\ 
   10  &   0.57 &    8.56 &    7.64 &    0.75 \\ 
\multicolumn{5}{c}{NGC~783}                \\
    1 &   0.73 &   8.42 &   7.57 &   0.88 \\
    3 &   0.57 &   8.56 &   7.90 &   0.71 \\
    4 &   0.82 &   8.46 &   7.65 &   0.81 \\
    6 &   0.77 &   8.46 &   7.74 &   0.79 \\
    7 &   0.84 &   8.34 &   7.61 &   0.91 \\
    8 &   0.58 &   8.57 &   7.91 &   0.72 \\
\multicolumn{5}{c}{NGC~2336}                 \\
   1a  &    0.88  &    8.47  &    7.61  &    0.81 \\
   1b  &    0.88  &    8.38  &    7.32  &    0.91 \\
    2  &    0.72  &    8.56  &    7.75  &    0.75 \\
    3  &    0.48  &    8.66  &    7.92  &    0.68 \\
    6  &    0.98  &    8.40  &    7.18  &    0.88 \\
   7a  &    0.65  &    8.54  &    7.69  &    0.76 \\
   7b  &    0.65  &    8.59  &    7.80  &    0.71 \\
    8  &    0.55  &    8.68  &    7.83  &    0.66 \\
    9  &    0.44  &    8.78  &    8.09  &    0.59 \\
  10a  &    0.50  &    8.63  &    7.90  &    0.68 \\
  10b  &    0.50  &    8.63  &    7.92  &    0.68 \\
   11  &    0.45  &    8.47  &    7.63  &    0.80 \\
   13  &    0.63  &    8.58  &    7.87  &    0.72 \\
  14a  &    0.97  &    8.43  &    7.43  &    0.88 \\
  14b  &    0.97  &    8.44  &    7.41  &    0.87 \\
  15b  &    0.79  &    8.47  &    7.43  &    0.84 \\
  16a  &    0.52  &    8.59  &    7.63  &    0.73 \\
  16b  &    0.52  &    8.64  &    7.76  &    0.69 \\
  17a  &    0.60  &    8.59  &    7.80  &    0.71 \\
  17b  &    0.60  &    8.57  &    7.92  &    0.72 \\
   19  &    0.56  &    8.62  &    7.66  &    0.71 \\
   20  &    0.62  &    8.57  &    7.81  &    0.72 \\
   21  &    0.57  &    8.52  &    7.94  &    0.74 \\
   22  &    0.76  &    8.54  &    7.69  &    0.76 \\
   23  &    0.74  &    8.52  &    7.68  &    0.78 \\
   24  &    0.66  &    8.44  &    7.57  &    0.84 \\
   25  &    0.65  &    8.48  &    7.59  &    0.82 \\
   26  &    0.53  &    8.50  &    7.69  &    0.78 \\
   27  &    0.55  &    8.47  &    7.61  &    0.81 \\
   28  &    0.86  &    8.50  &    7.64  &    0.79 \\
\multicolumn{5}{c}{NGC~6217}                \\
    1  &    0.48  &    8.65  &    7.89  &   0.68  \\
    3  &    0.47  &    8.46  &    7.65  &   0.83  \\
\multicolumn{5}{c}{NGC~7331}                \\
    1 &   0.49 &  8.44 &  7.59 &  0.85  \\
    3 &   0.17 &  8.41 &  7.39 &  0.69  \\
    4 &   0.45 &  8.53 &  7.71 &  0.76  \\
\multicolumn{5}{c}{NGC~7678}                \\
    1 &   0.90 &   8.45 &   7.42 &   0.87 \\
    2 &   0.59 &   8.51 &   7.67 &   0.78 \\
    3 &   0.56 &   8.30 &   6.98 &   0.80 \\
    4 &   0.56 &   8.55 &   7.67 &   0.76 \\
    5 &   0.56 &   8.52 &   7.92 &   0.75 \\
    6 &   0.49 &   8.55 &   7.71 &   0.74 \\
    7 &   0.67 &   8.49 &   7.58 &   0.80 \\
   8a &   0.59 &   8.53 &   7.66 &   0.77 \\
   8b &   0.59 &   8.54 &   7.74 &   0.76 \\
\hline
\end{tabular}\\
\end{center}
\end{table}

\setcounter{table}{4}
\begin{table}
\caption[]{Continued
}
\begin{center}
\begin{tabular}{ccccc} \hline \hline
 H\,{\sc ii} region                                 &
 R$_G$$^a$                                              &
12+log(O/H)$_{\rm NS}$                                  &
12+log(N/H)$_{\rm NS}$                                  &
t$_{\rm NS}$$^b$                                      \\ \hline
    9 &   0.82 &   8.51 &   7.75 &   0.78 \\
   10 &   0.38 &   8.52 &   7.63 &   0.77 \\
\hline
\end{tabular}\\
\end{center}
\begin{flushleft}
$^a$ in unit of isophotal radius $R_{25}$. \\
$^b$ $t_2$ in unit of 10$^4$ K. \\
\end{flushleft}
\end{table}

\begin{figure}
\resizebox{0.95\hsize}{!}{\includegraphics[angle=000]{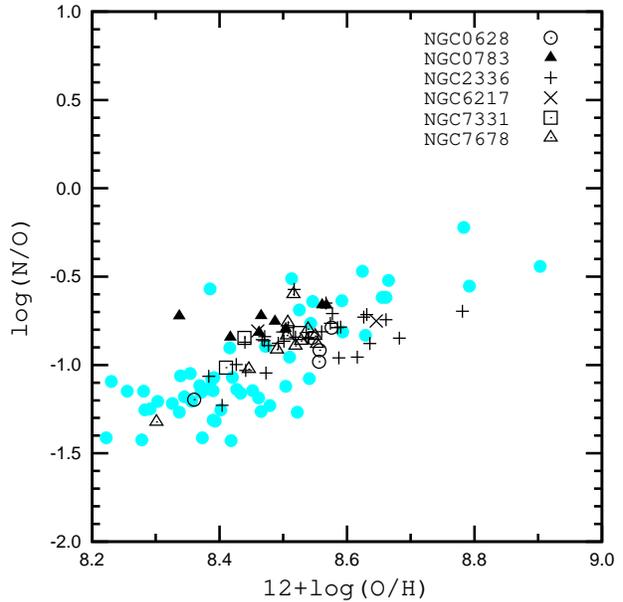}}
\caption{
The O/H--N/O diagram.
The different symbols marked in the legend
show abundances derived using the NS-calibration for our
sample of H\,{\sc ii} regions.
The filled circles show
$T_e$-based abundances in the sample of best-studied H\,{\sc ii}
regions in nearby galaxies (the compilation of data from
\citet{pilyugin2010}).
}
\label{figure:ohno}
\end{figure}

Both individual H\,{\sc ii} regions, excited by a single sourse
(single star or star cluster), and giant H\,{\sc ii} regions can be observed
in nearby galaxies. Giant H\,{\sc ii} regions can be composite nebulae which
contain a number of  H\,{\sc ii} regions with different physical properties,
all contributing to the global spectrum \citep{kennicutt1984}.
The ON and NS calibrations give oxygen and nitrogen abundances in composite
nebulae which agree with the mean luminosity-weighted abundances of their
components to within  $\sim$0.2 dex \citep{pilyugin2012}.

The calibrations are constructed under the assumption that the H\,{\sc ii}
regions are in the low-density regime. 
Since the sulphur lines [S\,{\sc ii}]$\lambda$6717 and [S\,{\sc ii}]$\lambda$6731
in most of our spectra are blended (the [S\,{\sc ii}]$\lambda$6717 and 
[S\,{\sc ii}]$\lambda$6731  were deblended in a several cases only and with
a low accuracy), we cannot  estimate the density sensitive ratio
[S\,{\sc ii}]$\lambda$6717/[S\,{\sc ii}]$\lambda$6731 
and verify the low-density regime in our objects.
Therefore we assume that the H\,{\sc ii} regions of our sample are all in the 
low-density regime, which is typical for the majority of extragalactic H\,{\sc ii} 
regions \citep{zaritsky1994,bresolin2005,gutierrez2010}.

In previous works, a little attention was paid to the radial
distribution of nitrogen
abundances in the disks of spiral galaxies, despite the fact that such
studies would have
several advantages \citep{thuan2010}.
First, since at $12+\log$(O/H) $\ga 8.3$,
secondary nitrogen becomes dominant and the nitrogen abundance increases at
a faster rate than the oxygen abundance \citep{henry2000},
the change in nitrogen
abundances with galactocentric distance should show a larger amplitude in
comparison to
oxygen abundances and, as a consequence, the gradient (and differences in
gradients
among galaxies)  should be easier to detect.
Furthermore, there is a time delay in the nitrogen production as compared
to oxygen production
\citep{maeder1992,vandenhoek1997,pagel1997book,pilyuginthuan2011}.
This provides an additional constraint on the chemical evolution of
galaxies.
For these reasons, not only
the radial distribution of oxygen abundances but also that of
nitrogen abundances are estimated here.

The resultant NS calibration abundances and electron temperatures
are given in Table~\ref{table:abun}. 
The oxygen and nitrogen abundances and electron temperature were 
derived  for H\,{\sc ii} regions where $R_3$, $N_2$, and $S_2$ lines 
are available. 
It should be noted that the NS-calibration relations  
have been derived using spectra of H\,{\sc ii} regions with well-measured electron 
temperatures as calibration datapoints. 
Therefore, the NS-calibration relations produce the abundances which are in agreement 
with the abundance scale defined by the classic $T_{\rm e}$  method.

\begin{figure*}
\resizebox{1.00\hsize}{!}{\includegraphics[angle=000]{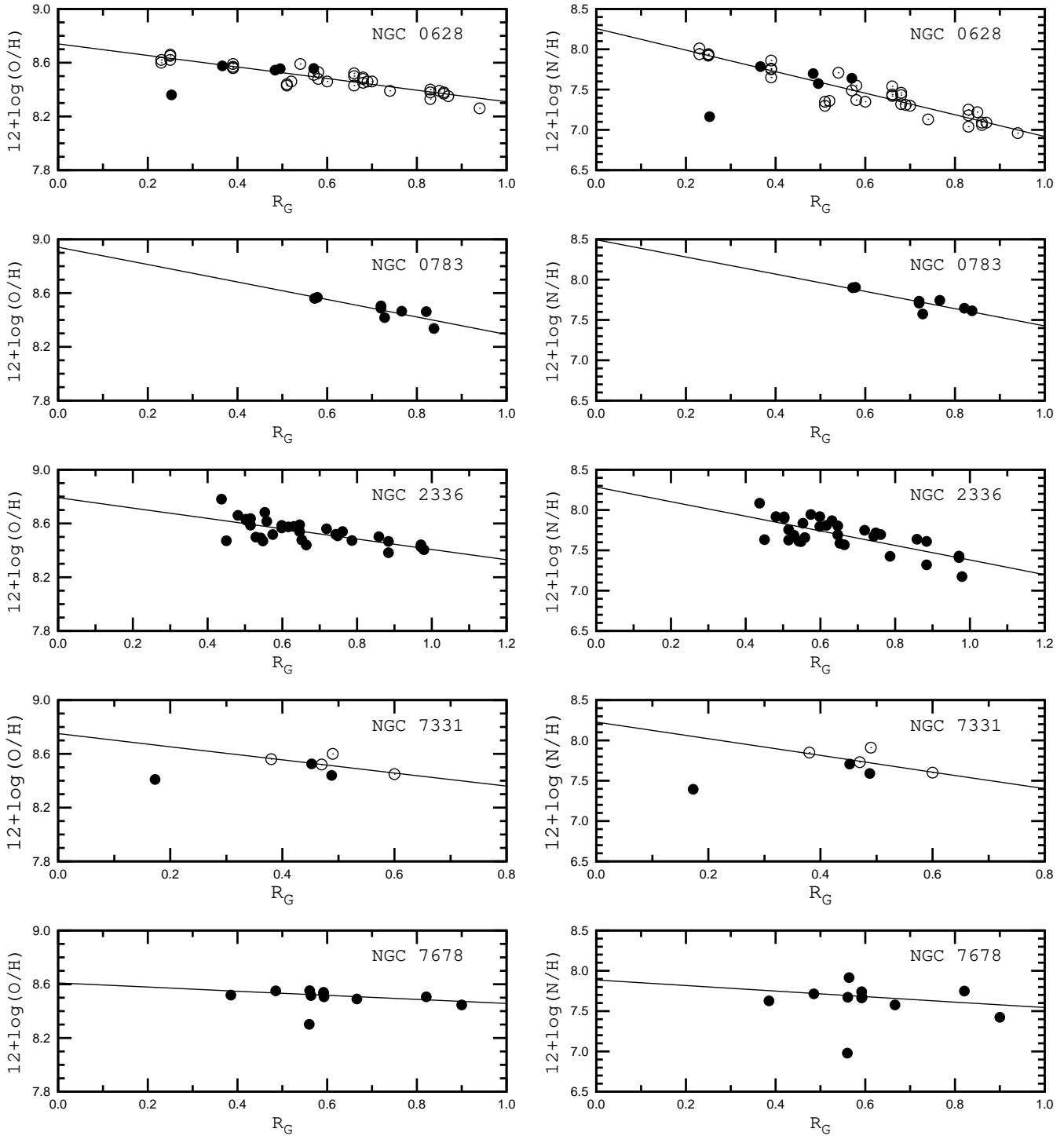}}
\caption{
Radial distributions of oxygen abundances (left column panels) and
nitrogen abundances
(right column panels) in the disks of galaxies.
Filled circles are abundances estimated from our observational data,
open circles are abundances based on the line measurements from the literature.
The solid lines are best fits to these datapoints.
}
\label{figure:grad}
\end{figure*}

The O/H--N/O diagram gives a possibility to test the validity of the determined abundances.
On the O/H--N/O diagram (Fig.~\ref{figure:ohno}) we compare abundances of 50 H\,{\sc ii} 
regions in six studied galaxies (see legend) with $T_e$-based abundances of the best studied H\,{\sc ii} 
regions (open circles) in nearby galaxies (data compiled from \citet{pilyugin2010}).
One can see from Fig.~\ref{figure:ohno}, that all points lie
within the spread of \citet{pilyugin2010}, i.e. the abundances in our
sample of H\,{\sc ii} regions  derived using
the NS-calibration occupy the same band in the  O/H - N/O diagram as the
$T_e$-based abundances  in the sample of best-studied H\,{\sc ii}
regions in nearby galaxies. 
This shows that our abundance estimations based on the NS-calibration method are realistic.

\subsection{Radial abundance gradients}

It is common practice \citep[e.g.][]{zaritsky1994,vanzee1998,pilyugin2004}
that the radial oxygen abundance distribution in the disk of spiral galaxy 
is fitted by the expression of the type:
\begin{equation}
12 + \log({\rm O/H})  = 12 + \log({\rm O/H})_0 + C_{\rm O/H} \times (R/R_{25}) ,
\label{equation:grado}
\end{equation} 
where 12 + log(O/H)$_{0}$ is the extrapolated central oxygen abundance,
$C_{\rm O/H}$ is the slope of the oxygen abundance gradient expressed in terms of dex/$R_{\rm 25}$,
and $R$/$R_{\rm 25}$ is the fractional radius (the galactocentric distance
normalized to the disk isophotal radius).
Analogously to the case of the oxygen abundance, the radial nitrogen abundance distribution in 
the disk of spiral galaxy is fitted by the equation of the type:
\begin{equation}
12 + \log({\rm N/H})  = 12 + \log({\rm N/H})_0 + C_{\rm N/H} \times (R/R_{25}) .
\label{equation:gradn}
\end{equation}

Fig.~\ref{figure:grad} shows the radial oxygen (left column of panels) and nitrogen 
(right column of panels) abundances distributions in the disk of spiral galaxies 
NGC~628, NGC~783, NGC~2336,  NGC~7331, and NGC~7678. The abundances derived from our 
spectral data are shown by the filled circles.

The emission line measurements in the spectra of H\,{\sc ii} regions in the disk 
of the galaxy NGC~628 previously were reported 
by \citet{mccall1985,ferguson1998,vanzee1998,bresolin1999} and in the disk of the 
galaxy NGC~7331 were given by \citet{mccall1985,ferguson1998}. 

As noted in Section~2.3, there are only two objects
in NGC~628 and NGC~7331 in common with previous spectroscopic observations. 
Nevertheless the abundances estimations in NGC~628 and NGC~7331 can be
directly compared with previously published ones. First we estimated the oxygen 
and nitrogen abundances through the NS calibration, using previously published line measurements. 
Further, we plotted these estimations on the diagram of the radial distributions of 
oxygen and nitrogen abundances in the disks of NGC~628 and NGC~7331 together 
with our abundances estimations.

The galactocentric distances were recomputed with the inclinations and position
angles adopted here (Table~\ref{table:sample}).

Fig.~\ref{figure:grad} shows a
good agreement in the radial abundances distribution in NGC~628 and NGC~7331
between H\,{\sc ii} regions observed in present paper (filled circles) and
H\,{\sc ii} regions observed previously (open circles) in these galaxies.

\begin{table}
\caption[]{\label{table:grad}
Parameters of radial distributions of oxygen and nitrogen abundances 
in the disks of galaxies.
}
\begin{center}
\begin{tabular}{ccccc} \hline \hline
galaxy                          &
O/H                             &
O/H                             &
N/H                             &
N/H                             \\ 
                                &
center$^a$                          &
gradient$^b$                        &
center$^a$                           &
gradient$^b$                         \\ \hline
NGC  628  & 8.74 $\pm$0.02 & -0.43 $\pm$0.03  & 8.26 $\pm$0.05  &  -1.34 $\pm$0.08  \\ 
NGC  783  & 8.94 $\pm$0.12 & -0.65 $\pm$0.16  & 8.49 $\pm$0.19  &  -1.07 $\pm$0.26  \\ 
NGC 2336  & 8.79 $\pm$0.05 & -0.38 $\pm$0.07  & 8.25 $\pm$0.10  &  -0.90 $\pm$0.15  \\ 
NGC 7331  & 8.75 $\pm$0.18 & -0.49 $\pm$0.36  & 8.23 $\pm$0.36  &  -1.03 $\pm$0.74  \\
NGC 7678  & 8.61 $\pm$0.03 & -0.15 $\pm$0.05  & 7.89 $\pm$0.17  &  -0.34 $\pm$0.27  \\ 
\hline
\end{tabular}\\
\end{center}
\begin{flushleft}
$^a$ in unit of 12+log(X/H). \\
$^b$ in unit of dex/$R_{25}$. \\
\end{flushleft}
\end{table}

\begin{figure}
\resizebox{1.00\hsize}{!}{\includegraphics[angle=000]{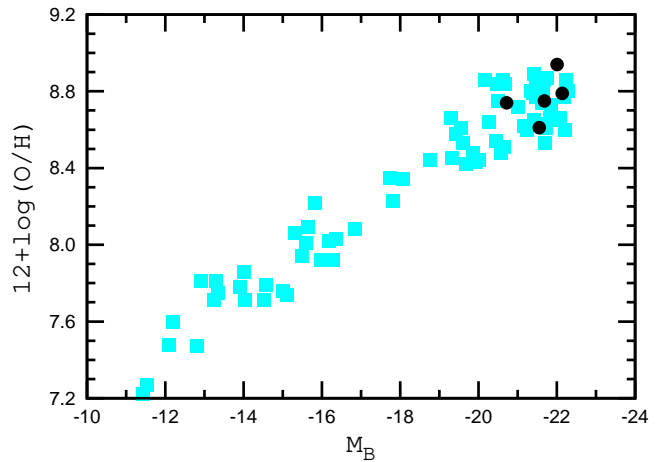}}
\caption{
The luminosity -- central metallicity diagram.
The dark (black) circles denote the central abundances in the
disks of spiral galaxies from the present sample.
The grey (blue) squares show the central abundances in the
disks of spiral galaxies and abundances in irregular
galaxies from \citet{pilyugin2007}.
(A colour version of this figure is available in the online version.)
}
\label{figure:mb}
\end{figure}

The numerical values of the coefficients in  Eq.(\ref{equation:grado}) ($C_{\rm O/H}$ and  12 + log(O/H)$_0$) and in Eq.(\ref{equation:gradn})  ($C_{\rm N/H}$ and  12 + log(N/H)$_0$) 
have been derived through the least squares method. Both our data and data from the literature are used. 
If there are datapoints which deviate significantly (more than 3$\sigma$) from the general trend, then these points were excluded in the derivation of the final relations. 
The obtained relations are presented on the Fig.~\ref{figure:grad} by the solid lines.
Computed parameters of radial distributions of oxygen and nitrogen abundances 
(the numerical values of the coefficients in  Eq.(\ref{equation:grado}) and Eq.(\ref{equation:gradn}))  
in the disks of galaxies NGC~628, NGC~783, NGC~2336, NGC~7331, and NGC~7678 
are listed in Table~\ref{table:grad}.

\subsection{Discussion}

Examination of Fig.~\ref{figure:grad} shows that the radial distributions 
of the oxygen and nitrogen 
abundances in the disks of considered galaxies are fitted well by 
the single linear expression within the optical isophotal radius.
The variation of nitrogen abundances with galactocentric distance shows a
larger amplitude in comparison to oxygen abundances and the differences in 
nitrogen abundance gradients among galaxies are larger than that 
in oxygen abundance gradients.

The luminosity -- central metallicity diagram for spiral and irregular 
galaxies have been constructed in \citet{pilyugin2007}. The abundances of 
H\,{\sc ii} regions in that paper and here have been estimated through 
the methods producing the abundances which are in agreement with the 
abundance scale defined by the classic $T_{\rm e}$ method.
Therefore the abundances obtained in these studies can be compared.
Fig.~\ref{figure:mb} shows the luminosity -- central metallicity diagram.
The grey (blue) squares show the central abundances in the
disks of spiral galaxies and abundances in irregular
galaxies from \citet{pilyugin2007}.
The dark (black) circles denote the central abundances in the
disks of spiral galaxies from the present sample.
The absolute blue magnitudes $M_B$ are taken from the {\sc leda} data base.
Inspection of Fig.~\ref{figure:mb} shows that our sample of galaxies
follows well the general trend in the luminosity -- central metallicity
diagram for spiral and irregular galaxies.

\section{Conclusions}

The  spectroscopic observations of  H\,{\sc ii} regions in six spiral galaxies (NGC~628, NGC~783, NGC~2336, NGC~6217, 
NGC~7331, and NGC~7678) obtained with the 6-meter telescope of 
Special Astrophysical Obsevatory (SAO) of the Russian Academy of Sciences
were carried out.

The oxygen and nitrogen abundances as well the electron temperatures in 
50 H\,{\sc ii} regions are estimated using the recent version of the 
strong line method (the NS-calibration).
The parameters of the radial distributions (the extrapolated central intercept 
value and the gradient) of the oxygen and nitrogen abundances 
in the disks of the NGC~628, NGC~783, NGC~2336 and NGC~7678 are obtained. 
The oxygen and nitrogen abundances and their gradients in the disks of 
spiral galaxies NGC~783, NGC~2336, and NGC~7678 are estimated for the first time. 
The abundances for the NGC~6217 are also found for the first time.

Galaxies from our sample follow well the general trend in the luminosity -- 
central metallicity diagram for spiral and irregular galaxies.

\section*{Acknowledgments}

We are grateful to the referee for his/her constructive comments.
L.S.P. acknowledges support from the Cosmomicrophysics project of
the National Academy of Sciences of Ukraine. A.S.G. is grateful to
A.Y.~Kniazev (South African Astronomical Observatory), to 
L.V.~Afanasiev and A.V.~Moiseev (Special Astrophysical
Observatory) for help and support during the observations in SAO
and for fruitful discussion, to A.V.~Zasov and B.P.~Artamonov (Sternberg
Astronomical Institute) for useful discussions. The authors
acknowledge the usage of the HyperLeda data base
(http://leda.univ-lyon1.fr). This study was supported in part by
the Russian Foundation for Basic Research (project nos.
08--02--01323, 10--02--91338, and 12--02--00827). Results based on 
observations collected with the 6-m telescope of the Special 
Astrophysical Observatory (SAO) of the Russian Academy of Sciences (RAS), 
operated under the financial support of the Science Department of 
Russia (registration number 01--43).

\end{document}